\documentstyle[aps,graphics]{revtex}

\input epsf

\begin{document}

\baselineskip=20.5pt
\def\fnote#1#2{\begingroup\def\thefootnote{#1}\footnote{
#2}
\addtocounter
{footnote}{-1}\endgroup}
\def\indt{\parindent2.5em}
\def\nd{\noindent}

\def\fo{\hbox{{1}\kern-.25em\hbox{l}}}  
\def\rf#1{$^{#1}$}
\def\bx{\Box}
\renewcommand{\theequation}{\arabic{equation}}

\vspace*{.2in}
\begin{center}
\large{\bf Photonic Hall Effect in ferrofluids: Theory and Experiments}\\
\vspace{26pt}
{D. Lacoste$^{a}$\fnote{*}{{\it e-mail: 
lacoste@dept.physics.upenn.edu}}}, F. Donatini$^{b}$, S. Neveu$^c$, J. A. Serughetti$^d$ and 
B. A. Van Tiggelen$^e$
\end{center}

\vskip 2mm
\begin{center}
$^a$~{Department of Physics,} \\
{University of Pennsylvania, Philadelphia, PA 19104, USA}\\
$^b$~{LRCCI, Universit\'e Jean-Monnet, } \\
{23 rue Docteur P. Michelon,
42023 St-Etienne Cedex 2, France}\\
$^c$~{LI2C, 4 place Jussieu, Case courrier 63, 75252 PARIS, France}\\
$^d$~{Universit\'e Claude Bernard Lyon I, } \\
{43 Blvd du 11 novembre 1918, 69622 Villeurbanne
CEDEX, France}\\
$^e$~{LPM2C, Maison des Magist\`eres Jean-Perrin,} \\
{CNRS-BP. 166, 38042 Grenoble Cedex 9, France}\\
\end{center}

\abstract
{An experimental and theoretical study on the Photonic Hall Effect (PHE) in liquid
and gelled samples of ferrofluids is presented. The ferrofluids are aqueous colloidal
suspensions of Fe\( _{2} \)CoO\( _{4} \) particles, which can be considered as
anisotropic and absorbing Rayleigh scatterers. 
 The PHE is found to be produced by the orientation of the magnetic
moments of the particles, as is also the case for the Faraday effect.
The dependence of the PHE with respect to the concentration of the scatterers,
the magnetic field and the polarization of the incident
light is measured in liquid and in gelled samples and is compared to
a simple model based on the use of a scattering matrix and the single
scattering approximation.}   

\vspace{28pt}
\vfill
\pagebreak

\setcounter{page}{1}

\section{Introduction }

Magneto-transverse light diffusion - more popularly
known as the ``Photonic Hall Effect" (PHE) - was theoretically predicted
five years ago by Van Tiggelen \cite{bart4}, 
and was experimentally confirmed 
one year later by Rikken \cite{nature}. 
This effect is analogous to the well-known electronic Hall effect, 
although the origin of the PHE is somewhat different:
 The PHE finds its origin in the Faraday effect,
present inside the dielectric
 scatterers, which slightly changes their
 scattering amplitude. 
Most experiments on the PHE reported so far used solid samples comprising paramagnetic
or diamagnetic particles, of size much larger than the wavelength (the Mie scattering
regime) \cite{nature}. The scatterers were embedded in
a medium having no magnetic properties, with a volume fraction such that
 multiple light scattering prevailed. A perturbational
formulation of the Mie scattering of a Faraday-active sphere has been developed
 to explain the origin of the PHE in single scattering \cite{josa}. 
A transport theory for light based on this formulation, could produce an 
estimate of the PHE in multiple light scattering, which agreed with
 experiments \cite{euro}. Experiments have checked the validity of this 
formulation by measuring the PHE as a function of the volume fraction of the scatterers,
of the wavelength or of the index of refraction even in the
 presence of absorption \cite{sabine2}.

These experiments can be difficult due to the smallness of the measured PHE.
It seems therefore natural to try to use other samples such as samples 
containing ferromagnetic particles, 
in which the magneto-optical effects are expected to be much larger than
 in paramagnetic or diamagnetic samples.
The giant magneto-optical properties of ferromagnetic compounds were
 explained by a microscopic theory due to Argyres, who derived the form of their polarizability
and conductivity tensors \cite{argyres}. This approach was
confirmed by Krinchik, whose classical model was able to explain both
the Faraday effect at optical frequency and the ferromagnetic Hall effect, which
is the classical Hall effect for the DC magneto-transverse electrical conductivity \cite{krinchik}.

\begin{figure}
{\par\centering \resizebox*{8cm}{8cm}{{\includegraphics{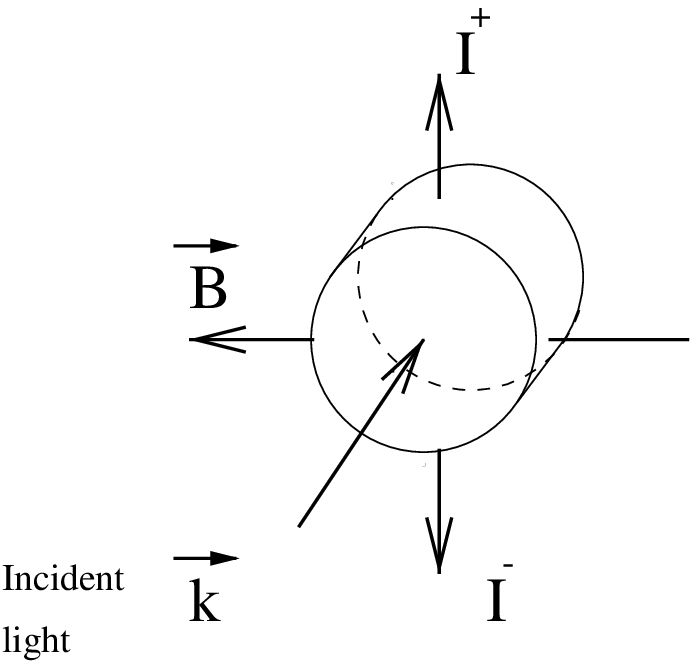}}} \par}
\caption{Sketch of the geometry of the PHE. The sample is shown as a cylinder. The
scattered intensities above (and below) the plane containing the incident light
direction and the magnetic field are called respectively  \protect\( I^{+}\protect \)
and  \protect\( I^{-}\protect \). The difference 
 \protect\( \Delta I({\mathbf{B}})\protect \)
is the photonic equivalent of a difference of potential in the electronic Hall effect.
\label{Fig:setup}}
\end{figure}

This paper deals with the PHE of liquid or gelled samples of ferrofluids, 
which are colloidal suspensions of ferromagnetic particles.
The experiments reported here
 are also the first experimental realization of the 
PHE in a medium comprising Rayleigh scatterers.
In these samples, the significant absorption 
 precludes the observation of multiple light scattering 
which does not exclude the PHE.
A sketch of the geometry is shown in Fig.~\ref{Fig:setup}.
 This geometry is called magneto-transverse (or Voigt geometry)
since the magnetic field and the incident light are perpendicular.
 These two perpendicular directions define a plane, and 
the total scattered intensity above this plane is called  \( I^{+} \) and the one
 below this plane
it is called \( I^{-} \). The PHE is a manifestation 
of an anisotropy in the
scattered light, which produces a net difference between  \( I^{+} \) and 
 \( I^{-} \),
linear in the magnetic field. The quantity  \( \eta  \), which is a
 measure of the PHE,
 is defined as the difference
between these intensities \( \Delta I({\mathbf{B}})= I^{+}({\mathbf{B}})-
 I^{-}({\mathbf{B}}) \) normalized with respect to
 the average of
the scattered intensities in the absence of any magnetic field denoted \( I_{0} \), 
\begin{equation}
\label{hall}
\eta =\frac{2(I^{+}({\mathbf{B}})-I^{-}({\mathbf{B}}))}{I^{+}({\mathbf{B}}=0)+I^{-}({\mathbf{B}}=0)}=\frac{\Delta I({\mathbf{B}})}{I_{0}}.
\end{equation}

The difference 
 \protect\( \Delta I({\mathbf{B}})\protect \)
is the photonic equivalent of the difference of potential in the electronic Hall effect.
In dilute ferrofluids, the ratio $\eta$ can be as high as $10^{-3}$ for a magnetic field of $100$Oe,
which is two or three orders of magnitude above the value measured in paramagnetic samples
 under the same conditions \cite{nature}.
 
This paper is organized as follows: Section \ref{Ferrofluids} recalls some general 
properties of ferrofluids, section \ref{magnetooptics}
 deals with their magneto-optical properties, section \ref{EHP:modele}
 presents a theoretical model for the 
PHE, and section \ref{experiments} contains the experimental study of the PHE
in liquid and gelled samples of ferrofluids.

\section{Ferrofluids}

\label{Ferrofluids} The stability of magnetic liquids can be destroyed by
several factors such as gravity, interaction between magnetic moments,
gradients in the applied magnetic field, and Van der Waals interactions, which
favor sedimentation or agglomeration. In practice, the stability of the suspension
is achieved by chemically grafting charged substituents or polymers on the particles,
in order to increase the repulsion between the particles, of ionic origin (in
aqueous ferrofluids), or of stearic origin (in organic ferrofluids). Because
of these stability requirements, the size of the particles in the suspension
is limited to a radius of a few nanometers \cite{rosensweig}. The ferrofluids
used in all experiments discussed in this paper, were synthesized by Neveu in the Laboratoire
de Physico-Chimie Inorganique in Jussieu using a coprecipitation technique first
devised by Prof. Massart (1981). 
The samples are suspensions of ferrite
particles Fe\( _{2} \)CoO\( _{4} \) dispersed in an aqueous ionic solution
of citrate.  

In such a suspension, each particle is a magnetic single-domain, of magnetic
moment
\begin{equation}
\label{moment}
\mu =4M_{s}\mu _{0}\pi a^{3}/3\approx 4,1 \cdot 10^{4}\mu _{B},
\end{equation}
 where \( M_{s}\simeq 422 \) kAm\( ^{-1} \) is the saturation magnetization
of the bulk material, and \( \mu _{B} \) is Bohr's magneton.  From this
average magnetic moment and the average distance between particles \( r \),
a dipolar interaction
energy \( E_{d}=(\mu _{0}/4\pi )\mu ^{2}/r^{3} \) can be estimated.
For the most concentrated of our samples of volume fraction \( f=1\% \), the
magnetic interactions are still negligible since at room temperature \( E_{d}/k_{B}T\simeq 0.17 \).
With such a small value, there should be essentially no chain structures
 in the magnetic liquid. The formation of chain structures in magnetic liquids was
 predicted long ago by De Gennes and Pincus \cite{degennes}, and was 
 studied more recently by Stevens and Grest using numerical simulations \cite{stevens}. 

In the absence of any magnetic field, the magnetic moments \( \mu  \) have
random orientation and the net average magnetization of the fluid is zero. When
a magnetic field is applied, there is a tendency for the particles to align
in the direction of the magnetic field. Langevin's theory of paramagnetic gases
applies to ferrofluids provided that the magnetic interaction are negligible.
The behavior of these magnetic fluids is known as superparamagnetism \cite{rosensweig}.
 The statistical average
 of the magnetic moments \( \overline{\mu }  \)
is oriented along the magnetic field and its amplitude is 
\begin{equation}
\label{langevin}
\overline{\mu }=\mu L(u),
\end{equation}
 where \( u=\mu B/k_{B}T \) for an applied magnetic field \( B \) and \( L(u) \)
is the Langevin function. It is recalled that \( L(u)=\coth u-1/u \), and \( L(u)\simeq u/3 \)
for \( u\ll 1 \). Quantitative
comparisons with experiments have shown that a polydispersity function should
be added to Eq.~(\ref{langevin}). Particles sizes obey a log-normal
 distribution characterized by an average radius \( a \)
and a variance.

The relaxation of the magnetization is known to have two possible mechanisms:
either it is caused by a rotation of the particles within the liquid, or by the 
rotation of the magnetic moments inside the particles with respect to their easy
magnetization axis. The first mechanism is characterized by the Brownian time relaxation
\cite{rosensweig}
\begin{equation}
\label{tauB}
\tau _{B}=\frac{3V'\nu}{k_{B}T},
\end{equation}
where \( V' \) is the hydrodynamic volume of the particle, and \( \nu \)
is the viscosity of the carrier liquid. The second mechanism is characterized
by the N\'eel relaxation time 
\[
\tau _{N}=\frac{1}{f_{0}}\exp (\frac{KV}{k_{B}T}),\]
 where \( K \) is the anisotropy constant, \( V \) the particle volume, and
\( f_{0} \) is a characteristic frequency of the order of \( 10^{9} \) Hz.
For Cobalt ferrofluids and for a magnetic field of frequency \( \omega _{H} \),
we have \( \tau _{B}\ll \omega _{H}^{-1}\ll \tau _{N} \). Therefore 
N\'eel relaxation
is negligible in these samples and the dynamic behavior of the fluids
 follows Brownian relaxation. The magnetic moments of the particles in Cobalt
ferrite samples remain fixed with respect to the particles 
as confirmed by studies of the linear birefringence
in these compounds \cite{neveu}.

\section{Magneto-optics of ferrofluids}

\label{magnetooptics}
\subsection{Optical properties in single scattering}

\begin{figure}
{\par\centering \resizebox*{8cm}{8cm}{\rotatebox{-90}{\includegraphics{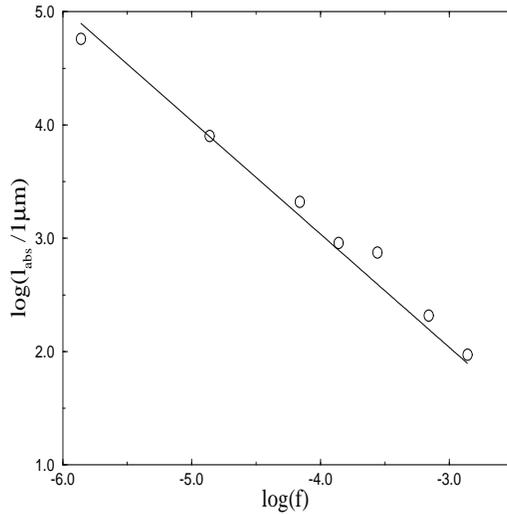}}} \par}
\caption{Logarithm of the absorption length log\protect\( _{10}(\ell _{abs}/1\protect \)\protect\( \mu \protect \)m\protect\( )\protect \)
as function of the logarithm of the volume fraction \protect\( f\protect \)
at the wavelength \protect\( \lambda _{0}=477\protect \)nm, for several aqueous
solutions of Fe\protect\( _{2}\protect \)CoO\protect\( _{4}\protect \).
 Points represent measurements and the line results from Lambert-Beer's law (the slope
is estimated from Rayleigh scattering theory for a monodisperse distribution
of particles of radius \protect\( a=6\protect \)nm using the index of refraction
of the material at this wavelength \protect\( 2.6-0.7i.\protect \)) \label{labs_f}}
\end{figure}

 Some aspects of the optical properties of ferrite Cobalt samples are now briefly
discussed. The relative dielectric constant \( \kappa _{0} \) in the absence
of magnetic field is obtained from the dielectric constant of the bulk material,
which equals \( 2.6-0.7i \)  at \( \lambda _{0}=477 \)nm according to Refs. \cite{krinchik2,martens}.
The average radius of the particles is estimated to be  \( a=6 \)nm.
This corresponds to Rayleigh scattering, since 
the average size parameter, \( x=2\pi an_{water}/\lambda _{0}\simeq 0.1 \),
 which is the ratio of the size of the particle
to the wavelength in the medium is smaller than one
as is also the case for the other size parameter  \( y=\sqrt{\kappa _{0}'}xn_{water}\simeq 0.2 \).
 For these Rayleigh absorbing particles, extinction
is dominated by absorption because of the behavior of the
scattering and absorption cross-section
as function of \( x \) \cite{bohren}.
For \( x\ll 1 \), and when
the imaginary part of the index of the scatterer is smaller than its real part
 (which is the case with these ferrofluids), the following relations hold

\[
Q_{abs}=4x\mathrm{Im} (\frac{\kappa _{0}-1}{\kappa _{0}+2})\, \, \gg \, \, Q_{scatt}=\frac{8x^{4}}{3}\left| \frac{\kappa _{0}-1}{\kappa _{0}+2}\right| ^{2}.\]
 The imaginary part of the index can therefore be determined from
the absorption spectra provided that the real part of the index is also known.
Figure (\ref{labs_f}) represents the absorption length as function of
the volume fraction on a logarithmic scale.
 The measurements were taken with a spectrometer, and the length of the sample 
was changed so that the absorbance, which is defined as the logarithm of the
 transmission, be of order unity.
The agreement with the linear behavior predicted from Lambert's law,
 with the slope computed from Rayleigh scattering theory, 
shows that our estimates concerning the index of refraction and the size of
the particles (assuming a monodisperse suspension) are basically correct.
The generalization to the case of polydisperse samples
 did not appear to be necessary for this estimate.

\subsection{Transport properties}

Typical values of the absorption length \( \ell _{abs} \) of the coherent beam
and of the transport mean free path \( \ell ^{*} \) in a sample of volume
fraction  \( f=0.14\% \), at the wavelength  \( \lambda _{0}=477 \)nm,
are  \( \ell _{abs}=80\mu  \)m (measured) and 
 \( \ell ^{*}=6 \)cm (theoretically estimated).
For these samples made of particles smaller than the wavelength,
the transport mean free path \( \ell ^{*} \) is very close to 
the elastic mean free path  \( \ell  \). A treatment of multiple light
scattering with absorption, becomes necessary when 
the elastic mean free path is smaller than the sample size. 
For all the samples discussed in this paper, the volume fraction is smaller than 
 \( f=0.1\% \), and the condition  \( \ell \gg L \) is fulfilled at this wavelength,
which means that multiple scattering should be negligible.

Because of the importance of absorption, 
 single scattering should therefore prevail.
It does not mean however that interference effects are absent, since 
scattered waves may still interfere as long as they have traveled a distance
shorter or comparable to the absorption length of the
coherent beam  \( \ell _{abs} \). This idea will be employed in
 the model of section \ref{EHP:modele}.

\subsection{Magneto-optical properties}

\label{Ssec:prop_magneto} 
\begin{figure}
{\par\centering \resizebox*{8cm}{8cm}{\rotatebox{-90}{\includegraphics{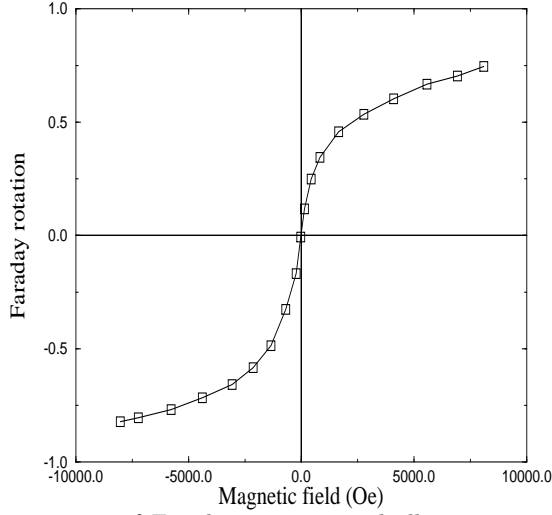}}} \par}
\caption{Normalized measurements of Faraday rotation and ellipticity in a sol-gel of 
Cobalt ferrite of volume fraction $0.04\%$.
The superposition of the normalized curves for the two effects is an indication 
of their common origin, both effects scaling with
 the Langevin function introduced in section \ref{Ferrofluids}. For the maximum field 
 \protect\( 8224\protect \)Oe, the Faraday rotation
is \protect\( 0.7^{\circ }\protect \) and the ellipticity \protect\( 1.91^{\circ }\protect \)
in a sample holder of length  \protect\( 2\protect \)mm.
The wavelength in this experiment was  \protect\( \lambda _{0}=633\protect \)nm,
and these measurements were taken in a static magnetic field. 
\label{Fig:faraday}}
\end{figure}

In the absence of an applied magnetic field, ferrofluids behave as a normal liquid, 
and no birefringence or dichroism, circular or  linear, are expected to be present.
The application of a magnetic field introduces magneto-optical anisotropy,
which has been the subject of many experimental studies \cite{davies,yusuf}.
In the longitudinal configuration, in which the wave vector of the incident light is
directed along the applied magnetic field, the eigenmodes of
the electric field are circularly polarized waves. The difference in speed and absorption
of these waves results in circular birefringence and dichroism, which is
 related to Faraday rotation and ellipticity.
The Faraday rotation and ellipticity are odd functions of the applied field
and are linear at small magnetic field as shown in Fig.~\ref{Fig:faraday}.
In this figure, only one curve is plotted, since the curves of the  
Faraday rotation and ellipticity
are both proportional to the same  Langevin function.
This was confirmed experimentally by measuring the sample magnetization,
 the Faraday rotation and ellipticity as a function of the magnetic field
at the  LRCCI laboratory with samples of 
Fe\( _{3} \)O\( _{4} \). All curves superimpose until
 volume fraction of a several percents \cite{jamon}.
This experiment confirms that the Faraday rotation and ellipticity in ferrofluids
have the same physical origin, which is the orientation of the magnetic
moments in the direction of the magnetic field. 

In the magneto-transverse or Voigt configuration, the eigenmodes of the electric field 
are linearly polarized waves. This implies linear birefringence and 
 dichroism, which are both even functions of the magnetic field. 
Below a critical field of about  \( 50 \)Oe for the Cobalt ferrite,
these effects are quadratic in the applied magnetic field. 
Above this critical field, the effects are linear but still of course
even functions of the applied magnetic field  \cite{daveze}. 
The linear birefringence and dichroism are also explained by Langevin theory
\cite{hasmonay}, as shown by the pioneering experiments of Bacri \emph{et al.} \cite{neveu,bacri}.
The standard Langevin model however, fails to describe the 
birefringence in spatially ordered samples or in samples in which the particles are
 not completely free to move. This can be the case for instance if the particles are gelled
 or binded to some substrate. The magneto-optical properties 
of gels were studied theoretically and experimentally 
in magnetic layered silica gels 
 in Refs.~\cite{bentivegna,visnovsky}. 
In random sol-gels of the type considered in this paper, the four magneto-optical
 effects have been measured \cite{donatini2}. 
These experiments have shown that the linear magneto-optical effects
 can be inhibited in sol-gels, whereas the circular magneto-optical effects
 are always present and essentially do not change
in the gels as compared to the liquids.

\section{Model for the PHE in ferrofluids}

\label{EHP:modele}
\subsection{Single scattering T matrix}
A simple model is presented to describe the Photonic Hall Effect
in liquid and gelled samples of ferrofluids. The samples are dilute and the predominant 
scattering corresponds to single scattering, which is described using a T matrix
formulation.
Our notations are as follows:  \( \hat{{\mathbf{k}}} \) and  \( \hat{{\mathbf{k}}}' \)
are the wave vectors of the incoming and outgoing light respectively and 
\( \hat{{\mathbf{B}}} \) is the direction of the applied field.
 With each of these three vectors, it is possible to associate a local
right-handed coordinate system, having its $z$-axis along the wave vector or
along the magnetic field. The $x$ and $y$ axis are chosen such that
 \( \hat{{\mathbf{B}}} \) and   \( \hat{{\mathbf{k}}} \) are
perpendicular, as is the case in the magneto-transverse geometry. 

The incoming beam can be regarded as a linear superposition of plane waves, each having
a transverse electric field
\begin{mathletters}
\begin{equation}
{\mathcal{E}}_{inc}={\mathbf{E}}_{inc}\exp\left[i(kz-\omega t+\delta ) \right],
\end{equation}
with
\begin{equation}
\label{elect}
{\mathbf{E}}_{inc}=E_1 \hat{{\mathbf{g}}}_{1}+E_2 \hat{{\mathbf{g}}}_{2}.
\end{equation}
\end{mathletters}
The wave vector of the light in the medium is denoted $k$, 
and the phase $\delta$ can be taken as a constant.
A convenient choice of polarization vectors is:
 \( \hat{{\mathbf{g}}}_{1}= \hat{{\mathbf{g}}}=\hat{{\mathbf{k}}}\times \hat{{\mathbf{k}}}'/
|\hat{{\mathbf{k}}}\times \hat{{\mathbf{k}}}'| \),
\( \hat{{\mathbf{g}}}_{2}=\hat{{\mathbf{k}}}\times \hat{{\mathbf{g}}}_{1} \).
The scattered beam can be decomposed in the polarization vectors 
\( \hat{{\mathbf{g}}}_{2}'=\hat{{\mathbf{k}}}'\times \hat{{\mathbf{g}}}_{1} \)
and \( \hat{{\mathbf{g}}}_{1}'=\hat{{\mathbf{g}}}_{1} \).
This choice of polarization vectors was introduced by Van De Hulst:
 one polarization vector is located in the scattering plane
 while the other one is perpendicular to it \cite{hulst}. This choice 
is not well defined in the particular cases of forward and backward scattering
 because there is no scattering plane in these cases, 
but these two particular configurations
 are not relevant for the PHE.
In the frame of the particle, in which quantities are denoted with a prime, 
the scattering matrix reads
\begin{equation}
\label{t:int}
\mathbf{t}'=\left( \begin{array}{ccc}
t_{0}' & it_{1}' & 0\\
-it_{1}' & t_{0}' & 0\\
0 & 0 & t_{0}'+t_{2}'
\end{array}\right).
\end{equation}
The complex-valued coefficients  \( t_{0}' \), \( t_{1}' \) and  \( t_{2}' \)
 describe a point-like scatterer in a magnetic field.

To show the relationship between the properties of a single particle and the macroscopic 
magneto-optical properties of an ensemble of particles, a statistical
 average of the magnetic moments of the particles has to be performed, 
as discussed in Ref. \cite{janssen}.
 When the coupling between the magnetic moments is neglected, 
and when a random distribution of scatterers is assumed,
the average T matrix has the same form as Eq.~(\ref{t:int}) with the following parameters
  \( t_{0}=t_{0}'+t_{2}' L(u)/u \), \( t_{1}=t_{1}'L(u) \)
and \( t_{2}=t_{2}'L_2(u)=t_{2}'(1-3L(u)/u) \), 
where \( u=\mu B/k_{B}T \) for an applied magnetic field \( B \), 
since
\begin{equation}
\label{t:moy}
\mathbf{t}=\overline{\mathbf{t}'}=\left( \begin{array}{ccc}
t_{0}'+t_{2}'L(u)/u & it_{1}'L(u) & 0\\
-it_{1}L(u) & t_{0}'+t_{2}'L(u)/u & 0\\
0 & 0 & t_{0}'+t_{2}'(1-2L(u)/u)
\end{array}\right).
\end{equation}
This result implies that the Faraday rotation associated with $t_1$,
 is proportional to the Langevin function
\( L(u) \), and that the linear birefringence is associated with the second Langevin function
\( L_2(u) \) contained in $t_2$, as confirmed by experiments.

The scattering matrix, which relates linearly the outgoing and the incoming electric
field needs now to be expressed in the basis of the polarization vectors introduced
above. This yields to the following result:
\begin{mathletters}
\begin{equation}
\label{t11}
S_{11}= t_{0}+t_{2} {\left( \hat{{\mathbf{B}}}\cdot \hat{{\mathbf{g}}} \right)}^2,
\end{equation}
  
\begin{equation}
\label{t12}
S_{12}= -t_{2} \frac{ \left(\hat{{\mathbf{B}}}\cdot \hat{{\mathbf{k}}}' \right)
 \left(\hat{{\mathbf{B}}}\cdot \hat{{\mathbf{g}}} \right) }{|\hat{{\mathbf{k}}}\times \hat{{\mathbf{k}}}'|},
\end{equation}

\begin{equation}
\label{t21}
S_{21}=-it_{1} \hat{{\mathbf{B}}}\cdot \hat{{\mathbf{k}}}' 
 -t_{2} \frac{ \left( \hat{{\mathbf{B}}}\cdot \hat{{\mathbf{k}}}' \right) \,
 \left(\hat{{\mathbf{B}}}\cdot \hat{{\mathbf{g}}} \right) \,
 \left(\hat{{\mathbf{k}}}\cdot \hat{{\mathbf{k}}}' \right) }
{|\hat{{\mathbf{k}}}\times \hat{{\mathbf{k}}}'|},
\end{equation}

\begin{equation}
\label{t22}
S_{22}=t_{0} \hat{{\mathbf{k}}}\cdot \hat{{\mathbf{k}}}' + t_{2} \frac{ 
\left(\hat{{\mathbf{k}}}\cdot \hat{{\mathbf{k}}}' \right)
{\left( \hat{{\mathbf{B}}}\cdot \hat{{\mathbf{k}}}' \right)}^2 }
 {{|\hat{{\mathbf{k}}}\times \hat{{\mathbf{k}}}'|}^{2}}
-it_{1} \hat{{\mathbf{B}}}\cdot (\hat{{\mathbf{k}}}\times \hat{{\mathbf{k}}}').
\end{equation}
\end{mathletters}

\begin{figure}
{\par\centering \includegraphics{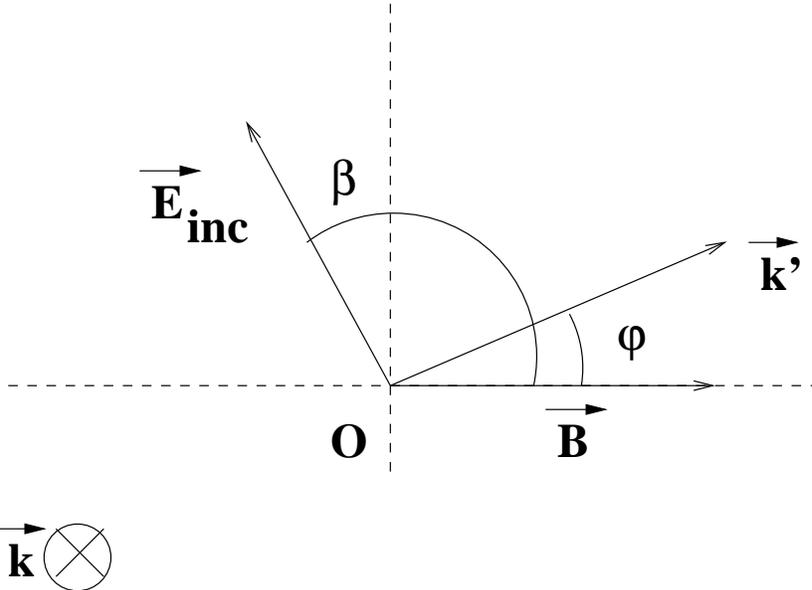} \par}
\caption{Sketch of the geometry in the frame of the laboratory.
The vector  \protect\( {\mathbf{B}}\protect \)
defines the axis  \protect\( x\protect \), the magneto-transverse direction
is along the axis 
\protect\( y\protect \), and the incident wave vector is along
 \protect\( z\protect \). The angle $\beta$ characterizes the state of linear
polarization of the incident electric field. 
 \label{schema:ff}}
\end{figure}

These expressions are valid only for the magneto-transverse geometry but 
at any order in the magnetic field.
They can be simplified by introducing a coordinate system
in which the wave vector of the scattered wave
\( \hat{{\mathbf{k}}}' \) makes an angle \( \theta  \)
in the scattering plane with respect to the incoming direction 
and an azimuthal angle \( \varphi  \) 
with respect to  \( \hat{{\mathbf{B}}} \) as shown in figure \ref{schema:ff}.
With this choice,
\[
\hat{{\mathbf{B}}}=\left( \begin{array}{c}
1 \\
0 \\
0
\end{array}\right) ,\, \, \, \, \, \, \, \hat{{\mathbf{k}}}=\left( \begin{array}{c}
0\\
0\\
1
\end{array}\right) ,\, \, \, \, \, \, \, \hat{{\mathbf{k}}}'=\left( \begin{array}{c}
\sin \theta \cos \varphi \\
\sin \theta \sin \varphi \\
\cos \theta 
\end{array}\right) .\]

\subsection{PHE for unpolarized incident light}

The scattering matrix of Eqs. (\ref{t11}-\ref{t22}) describes single
 scattering. This approach is not sufficient to account for the PHE in ferrofluids,
 because the PHE of a single magnetic Rayleigh scatterer is zero as was found in Ref.~\cite{josa}. 
It is therefore necessary to take into account interferences among different 
single scattering events, in order to explain the observed non-zero value of the PHE.
To this end, one can start with the scattering matrix
 of a set of particles at position  \( {\mathbf{r}}_{i} \) 
and of moment  \( \mu _{i} \), 
which can be written  
\begin{equation}
\label{Tmatrix}
\left< {\mathbf{S}} \right>=\sum _{i}{\mathbf{S}}(\mu _{i})\, \left| {\mathbf{r}}_{i}\right\rangle \, \left\langle {\mathbf{r}}_{i}\right| .
\end{equation}
The summation can be performed with the following approximation.
Because of the significant absorption in the samples,
it is meaningful to neglect all interferences effects outside a range of the order
of the absorption length of the coherent beam \( \ell _{abs} \). In a medium of
 size \( \ell _{abs} \),
which is assumed to be spherical, interferences can be
taken into account exactly in single scattering  
since  \( \ell ^{*}\gg \ell _{abs} \). In the basis of Van De Hulst for the
polarization, the scattering matrix has the following form 
\begin{equation}
\label{mat}
\left< S_{ij} \right> \approx n\int _{0}^{\ell _{abs}}
e^{i({\mathbf{k}}-{\mathbf{k}}')\cdot
 {\mathbf{r}}_i}
\, S_{ij} \, d^{3}{\mathbf{r}}_i=f\, \left( \frac{\ell _{abs}}{a}\right)
 ^{3}G(v)\, S_{ij},
\end{equation}
where $n$ is the number of particles per unit volume, \( G(v)=3(\sin v-v\cos v)/v^{3} \) is the phase function for Rayleigh-Gans
scattering and 
\( v=2k\ell _{abs}\sin (\theta /2) \). 
Relation (\ref{mat}) expresses a well known result in 
optics: the proportionality of the scattering matrix of a Rayleigh-Gans
to the scattering matrix of a Rayleigh  scatterer \cite{hulst}. 
From the scattering matrix, the scattering cross-section can be computed
 for the given states of incident
 and outgoing polarizations. 
The scattering cross-section, averaged over  
incident and outgoing
polarization, reads:
\begin{equation}
\label{phe}
\left\langle \frac{d\sigma }{d\Omega }\right\rangle =
\frac{1}{4} f^2 \left( \frac{\ell _{abs}}{a}\right)
 ^{6} G^2 \left( v \right) \, \sum_{i,j=1,2} |S_{ij}|^2.
\end{equation}
To compute the PHE \( \eta  \) defined in Eq.~(\ref{hall}), 
it is necessary to integrate the scattering cross-section
with respect to the direction \( \hat{{\mathbf{k}}}' \)
including a magneto-transverse projection factor 
 \( \hat{{\mathbf{B}}}\cdot (\hat{{\mathbf{k}}}\times \hat{{\mathbf{k}}}') \).
This factor allows to calculate the light flux
projected onto the magneto-transverse direction
 \( \hat{{\mathbf{B}}}\times \hat{{\mathbf{k}}} \), which is
the direction of the normal of the detector in the experiment.
After this integration, only terms odd in the magnetic field remain.
For unpolarized incident light, the PHE finally reads 
\begin{equation}
\label{integral}
\eta_{unpol} =\pi \gamma _{1}\frac{\int ^{\pi }_{0}d\theta \sin ^{3} \theta 
\cos \theta \, G^2 \left[ v(\theta )\right] }{\int ^{\pi }_{0}d\theta
 \sin ^{2} \theta (1+\cos ^2 \theta ) \, G ^2 \left[ v(\theta )\right] },
\end{equation}
with  
\begin{equation}
\label{mu1}
\gamma _{1}=\frac{\Im m(t_{0}t^{*}_{1})}{|t_{0}|^{2}}.
\end{equation}
 The integrals in Eq. (\ref{integral}) can be calculated in the limit 
\( k\ell _{abs}\gg 1 \), which gives
\begin{equation}
\label{etafinal}
\eta_{unpol} =\frac{3\gamma_{1}}{2}\frac{\ln (k\ell _{abs})}{k\ell _{abs}}.
\end{equation}
It is interesting to note that when the
size parameter of the particles goes to zero  \( x\rightarrow 0 \),
 the parameters  \( \gamma _{1} \) and \( \eta_{unpol}  \)
are independent of the size of the particles. This makes this limit
insensitive to polydispersity, which is always difficult to estimate in 
this kind of experiment.

\subsection{PHE for polarized incident light}

Polarized incident light may be either circular or linear. 
A state of linear polarization is described by 
\begin{equation}
E_1=\cos\beta, \,\,\,\,\,\,\, E_2=\sin\beta,
\end{equation}
in Eq.~(\ref{elect}).
A sketch of the geometry is shown for this case in Figure \ref{schema:ff}.
When  \( \beta =0 \), the electric field of the incident light is 
parallel to the applied magnetic field.

Contrary to the case of unpolarized PHE, several terms present in the scattering matrix
now contribute.
After having carried out the integration over the outgoing wave vector and
 the average with respect to the 
outgoing polarization, the
numerator  \( \Delta I({\mathbf{B}}) \) and the denominator  \( I_{0} \) of
the PHE take the form  
\begin{eqnarray*}
\Delta I({\mathbf{B}}) & = &  \pi  A
\left[ -\frac{1}{2}\mathrm{Im}(t_{2}t_1^{*})
\cos(2\beta) + 2 \mathrm{Im}(t_{0}t_1^{*}) \sin ^{2} \beta \right]
\int_0^{\pi} d\theta \sin^3 \theta \cos\theta \, G^2 \left[ v(\theta) \right]
,\\
I_{0} & = &  2  A |t_0|^2
\int_0^{\pi} d\theta
\left[ \cos^2 \beta + \cos^2 \theta \sin ^2 \beta  \right]
 \sin^2\theta \,  G^2\left[ v(\theta) \right],
\end{eqnarray*}
where $A$ is a known constant of proportionality.
In these expressions, the integrals over the scattering angle 
$\theta$ can be done in the limit \( k\ell _{abs}\gg 1 \). For the last equation, this gives 
\begin{equation}
I_0= 2 A |t_0|^2
 \int_0^{\pi} d\theta \sin^2\theta \, G^2\left[ v(\theta )\right],
\end{equation}
which means that the denominator  \( I_{0} \) of the PHE is independent
of the angle  \( \beta  \) characterizing the state of linear incident polarization.
The PHE for polarized incident light now reads:
\begin{equation}
\label{polar:ff}
\eta (\beta )=2\eta _{unpol} \left[ \sin ^{2} \beta -\frac{\gamma_2}{4 \gamma_1}
 \cos(2\beta) \right],
\end{equation}
where  \( \eta _{unpol} \) is the value of the PHE obtained from Eq.~(\ref{etafinal})
for unpolarized incident light, and $\gamma_2$ has been defined in the same way as in 
Eq.~(\ref{mu1}) by
\begin{equation}
\label{mu2}
\gamma _{2}=\frac{\mathrm{Im} (t_{2}t^{*}_{1})}{|t_{0}|^{2}}.
\end{equation}
Averaging over  \( \beta  \)
in  Eq.~(\ref{polar:ff}) reproduces the value  \( \eta _{unpol} \).
Two particular cases of Eq.~(\ref{polar:ff}) will be of interest:
when the linear magnetic birefringence is negligible, one has 
\begin{equation}
\label{case1}
\eta (\beta )=2\eta _{unpol} \sin ^{2} \beta.
\end{equation}
On the other hand, for a dominant contribution of the birefringence, 
one gets
\begin{equation}
\label{case2}
\eta (\beta )=- \eta _{unpol} \frac{\gamma_2}{2 \gamma_1}
 \cos(2\beta)= - \frac{3\gamma_2}{4} \frac{\ln (k\ell _{abs})}{k\ell _{abs}}
 \cos(2\beta).
\end{equation}

The dependence of the PHE on the state of circular polarization can be done very
similarly. A state of circular polarization can be written  
\[
E_1=\frac{1}{\sqrt2}, \,\,\,\,\,\,\, E_2=\frac{\pm i}{\sqrt2}.
\]
 One finds that the PHE is independent of the state of circular polarization
left or right of the incident light 
\begin{equation}
\label{circul}
\eta ^{\pm }=\eta _{unpol}.
\end{equation}
This value corresponds to the case of linear polarization
at the angle \( \beta =\pi /4 \), or to the case of
 unpolarized incident light.

\section{Experimental results}
\label{experiments}
\subsection{Dependence of the PHE as function of the field and concentration}

The light intensities  \( I^{+} \) and \( I^{-} \),
represented in Fig.~\ref{Fig:setup} travel trough optical fibers 
from the sample to the detectors which are photo-diodes or photo-multipliers.
The difference \( \Delta I({\mathbf{B}}) \) is measured using a lock-in.
More details on the experimental setup can be found in Ref.~\cite{nature}.
The measurement of the phase difference of \( \Delta I({\mathbf{B}}) \) with
respect to the magnetic field defines the sign of the PHE. This phase
for a true signal should be therefore either $0$ or $\pi$ radians.
The sign of the PHE is a very important feature which has been seen
to be very sensitive to several experimental parameters (Verdet constant, index of
refraction, concentration, polarization..).

\begin{figure}
{\par\centering \resizebox*{8cm}{8cm}{\includegraphics{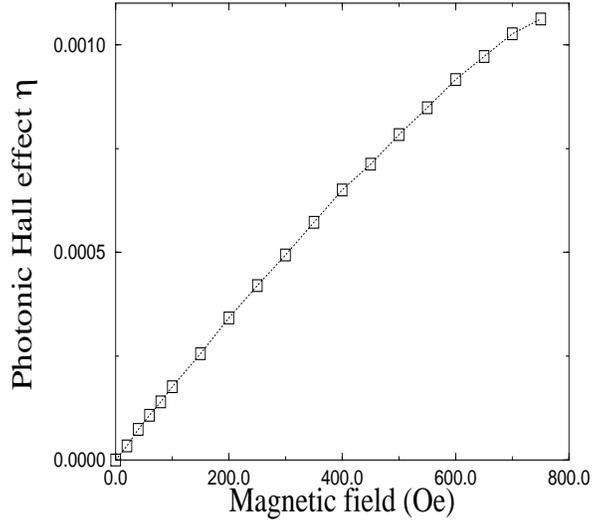}} \par}
{\par\centering \resizebox*{8cm}{8cm}{\includegraphics{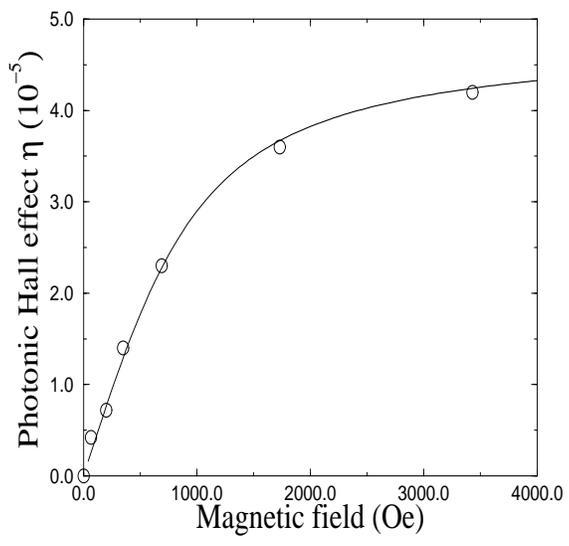}} \par}
\caption{(a) The squares represent measurements of the PHE \protect\( \eta \protect \)
at low field, as function of the applied magnetic field. The curve only connects
the experimental points. It can be noted that the PHE is linear in magnetic fields
smaller than about a hundred of Oe.
This experiment was done with the same sample than in Fig.~(\ref{Fig:faraday}).
 (b) Same experiment with similar samples but with higher magnetic field,
so that the saturation of the PHE is now clearly visible. The circles are 
experimental points and the line is a fit using the 
Langevin function. \label{Fig:hall}}
\end{figure}

Our first experimental study addresses the dependence of the PHE as a function
of the applied magnetic field. As shown in Fig.~\ref{Fig:hall},
the PHE is linear in the applied magnetic field at low field. 
In ferrofluids, the deviation away from a linear behavior sets in at fields higher
than a few hundreds of Oe. Experiments done at higher field show the same
saturation which is observed in figure
 \ref{Fig:faraday} for the Faraday effect.
 This aggrees with the prediction of Eq.~(\ref{mu1}) and Eq.~(\ref{etafinal}),
and confirms that the PHE is
produced by the orientation of the magnetic moments according
to the Langevin model. 
Unless specified otherwise, the rest of the measurements of the PHE
was done in the linear regime for the PHE.

\begin{figure}
{\par\centering \resizebox*{8cm}{8cm}{\rotatebox{-90}{\includegraphics{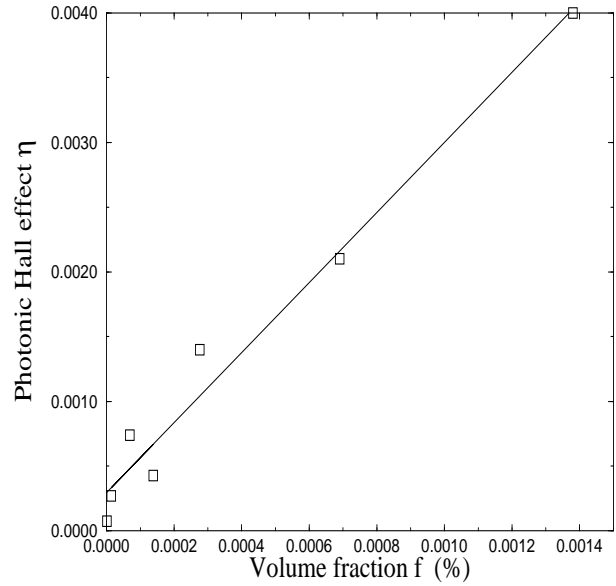}}} \par}
\caption{Points are experimental measurements of PHE  \protect\( \eta \protect \)
as function of the volume fraction  \protect\( f\protect \).
Samples are gels made with \protect\( 1\%\protect \)
of gelatine in volume and Cobalt ferrofluid. 
A linear fit gives the slope per unit of magnetic field 
 \protect\( d^{2}\eta /dBdf=3 \cdot 10^{-3}T^{-1}.\protect \)
\label{vol_f}}
\end{figure}

The variation of the PHE with volume fraction  \( f \), is 
shown in Fig.~\ref{vol_f} for a gelled sample, which
  was prepared with about  \( 1\% \) of gelatine
in volume. The experiment was repeated with liquid samples of
ferrofluids. In both cases, a linear behavior with respect to 
the volume fraction was found. The dispersion of the experimental
points in Fig.~\ref{vol_f} around the linear behavior 
is believed to be due to the evaporation of water
in the gelled samples, and was absent in the
 experiment with liquid samples.
The approximation used in Eq.~(\ref{etafinal}) is
 valid since \( k\ell _{abs}\simeq 1.6\cdot 10^{3} \)
for the most concentrated of the samples used of volume fraction  \( f=0.14\% \).
Since the absorption length \( \ell _{abs} \) is inversely proportional
to the volume fraction of the scatterers, Eq.~(\ref{etafinal}) predicts
 a linear dependence of  \( \eta  \) with respect to the volume fraction 
\( f \). This must be caused by a difference  \( \Delta I({\mathbf{B}}) \)
 quadratic in \( f^{2} \) while the scattered intensity  
\( I_{0} \) is proportional to \( f \). A quadratic behavior for  \( f^{2} \)
in  \( \Delta I({\mathbf{B}}) \) means that the PHE involves more than
one particle, because the probability of scattering for a photon is  
 proportional to  \( f \). These experiments confirm our previous statement that
 a single Rayleigh scatterer is not able to
 generate the PHE but that more than one Rayleigh
 scatterers or one scatterer of finite size
 is necessary to get a non-zero value of the PHE \cite{josa}.

The PHE per unit magnetic field in the experiment of Fig.~\ref{vol_f}
is estimated to be  \( d^{2}\eta /dBdf=3.10^{-3}T^{-1} \).
In order to compare this value with a theoretical estimate, it is necessary
to evaluate the coefficients of the scattering matrix.
For a low applied magnetic field, these coefficients can be determined
 from the anisotropy in the index of refraction \cite{bart}. 
The dielectric constant of the sample is assumed to have the following form
at low applied magnetic field:
\begin{equation}
\epsilon_{ij}=\epsilon_0 \delta_{ij}+i\epsilon_F \epsilon_{ijk}\hat{B}_k.
\end{equation}
Using our measurements of the Faraday rotation and ellipticity
in this sample, we estimate the antisymmetric part of
the dielectric constant \( \varepsilon _{F} \) to be
\( d\varepsilon _{F}/dB\simeq (-0.5+1.5i)\cdot 10^{-4}\, T^{-1}. \)
With this value, one can calculate the antisymmetric part of the relative 
dielectric constant \( \kappa _{F} \) of a single particle. The parameter
 \( \gamma _{1} \) can then be obtained from the relation 
\[
\gamma _{1}=\frac{\mathrm{Im} (t_{0}t^{*}_{1})}{|t_{0}|^{2}}=-3\mathrm{Im}
 \left[ \frac{\kappa _{F}}{(\kappa _{0}-1)(\kappa _{0}+2)}\right].\]
This yields \( d^{2}\eta /dBdf\simeq 1,2 \cdot 10^{-3}\, T^{-1}, \)
which is to be compared with the slope of Fig.~\ref{vol_f}
 \( d^{2}\eta /dBdf\simeq 3\cdot 10^{-3}\, T^{-1}. \)
Because of experimental uncertainties due to the lack of stability
of the gel, this can be considered as a fair agreement.
The other experiments with gels which are reported in this paper, 
were done with sol-gels which are stable.

\subsection{Experimental study of the polarization of the PHE}

The dependence of the PHE on the volume fraction is an important
test of the model.
However a lot of physical and optical properties 
depend on the volume fraction, which makes the interpretation
of this kind of experiments difficult.
The polarization is a much more robust quantity to consider, since it 
should be independent of many experimental conditions.
This section adresses
 the dependence of the PHE on the
incident polarization for liquid and gelled samples of ferrofluids.
 The polarization 
of the scattered light is still averaged by the detector, which is 
not sensitive to the polarization of the light,
as was checked experimentally. 
The opposite experiment, consisting of sending unpolarized
 light and measuring
the polarization of the outcoming light, is equally interesting,
but has not been carried out, because it is more difficult to
realize.

\begin{figure}
{\par\centering \resizebox*{8cm}{8cm}{\rotatebox{-90}{\includegraphics{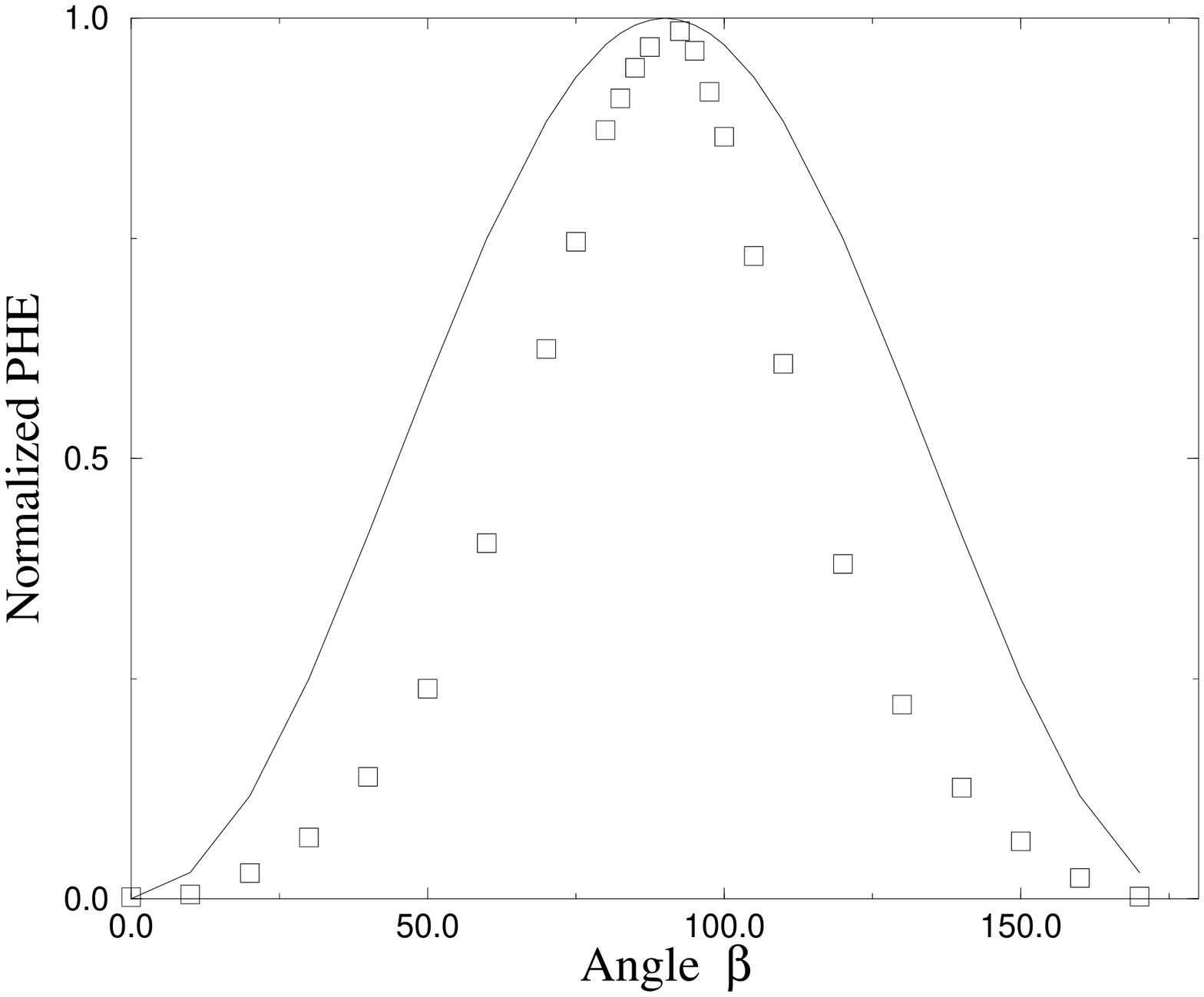}}} \par}
{\par\centering \resizebox*{8cm}{8cm}{\rotatebox{-90}{\includegraphics{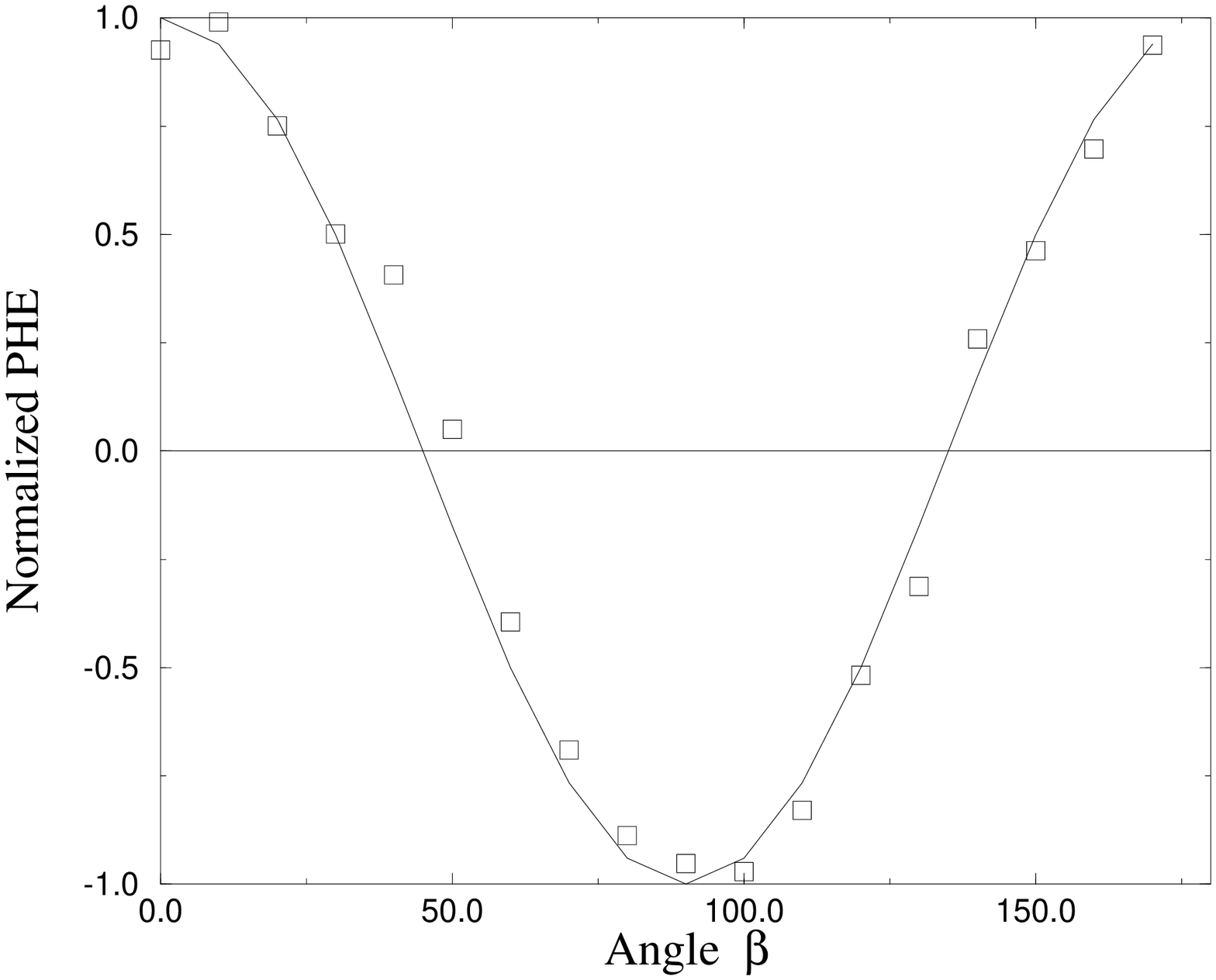}}} \par}
\caption{(a)  Normalized PHE for a  sol-gel sample of
 ferrofluid as function of the angle  \protect\( \beta \protect \)
 of the incident linear polarization. Points are experimental
and the curve is a prediction from Eq.~(\ref{case1}).
A magnetic field of 
\protect\( 250\protect \)Oe and frequency \protect\( 560\protect \)Hz 
was applied on the same sample studied in Fig.~\ref{Fig:faraday}b. (b) 
Normalized PHE for liquid ferrofluid as 
 function of  \protect\( \beta \protect \), together with the prediction
of Eq.~(\ref{case2}).
A magnetic field of  \protect\( 40\protect \)Oe and frequency \protect\( 20\protect \)Hz,
was applied on a sample of
 volume fraction  \protect\( 0.025\protect \)\%
of aqueous Cobalt ferrite ferrofluid. 
\label{polarisation}}
\end{figure}

The first experiment deals with sol-gels of ferrofluids.
The gels were produced in the group of J. A. 
Serughetti from D\'epartement de physique
 des Mat\'eriaux de l'Universit\'e
Claude Bernard in Villeurbanne. 
The gels were made from the isotropic liquid state with no magnetic
field present during the gelation. 
During the measurement of the PHE, 
a magnetic field of  \( 250 \)Oe and of frequency  \( 560 \)Hz
was applied.
 In Fig.~\ref{polarisation}a
 the normalized PHE is shown as a function of \( \beta  \), together with the prediction
from Eq.~(\ref{case1}), when magnetic linear birefringence, 
the second term in Eq.~(\ref{polar:ff}),
 is assumed to be absent.
This assumption is consistant with the observed weak sensitivity of this curve 
with respect to the amplitude of the applied 
magnetic field. We have checked that the dependence of
 the PHE on \( \beta  \)
for linearly polarized incident light comes in only from the numerator of 
 \( \eta  \) in Eq.~(\ref{hall}), as no dependence was found
in the incoherent background \( I_{0} \) as predicted by the
 model of the previous section.
The magnetic birefringence was assumed to be absent in the comparison with our experiments 
since the it should be inhibited in the gel.
Although the general features of the curve are captured by the model,
 there remains a  clear discrepancy between experiments and 
 theory. This discrepancy 
follows a  \( \cos \beta ^{2}\sin ^{2}\beta  \) law,
which probably indicates that this point might be explained by taking
 into account higher scattering orders, such as double scattering for 
instance.

Figure \ref{polarisation}b shows measurements of the PHE as
function of the angle \( \beta  \) 
together with the theoretical prediction 
of Eq.~(\ref{case2}).
The sample is a liquid Cobalt ferrofluid, of volume fraction
 \protect\( 0.025\protect \)\%, and the applied 
magnetic field is  \protect\( 40\protect \)Oe.
It can be first noted that the observed dependence of the PHE is
completely different from the previous case:  the symmetry of the
curve with respect to the zero axis is different. There is now
 a range of values of \( \beta  \) for which the PHE changes sign, and 
the effect no longer vanishes at $\beta=0$, but is maximum. 
Contrary to the case of gelled samples, the experimental result
 is now compared to the theoretical model
for dominant linear magnetic birefringence. 
In these ferrofluids,
the particles are indeed free to reorganize themselves 
and acquire shape anisotropy,
which results in a linear magnetic birefringence generally 
 larger than the Faraday effect. 
In the same experimental conditions of figure (\ref{polarisation}b) and 
in a sample length of $3$mm,  
the Faraday rotation was measured to be $0.1^{\circ }$,
 to be compared to
$1.7^{\circ }$ for the linear magnetic birefringence.
The magnetic linear birefringence corresponds to the term  
 $\gamma_2$ in Eq.~(\ref{case2}), which is an odd function
 of the magnetic field very much like $\gamma_1$ in 
the unpolarized case.
 At low magnetic field,  $\gamma_2$ is proportional to ${\mathbf{B}}^3$, 
and should therefore generate an oscillation at the
 modulation frequency of the magnetic field.
This time, a very good agreement with the predicted
 law in \( -\cos (2\beta ) \) is found.

The experiment was repeated with circularly polarized incident light, in
liquids and in gels, and fully
confirmed the result of Eq.~(\ref{circul}).
There is virtually no measurable difference between  \( \eta ^{+} \) et \( \eta ^{-} \).
 The value of \( \eta ^{+} \) corresponds indeed to
the value of the PHE with linear polarization for an angle
\( \beta =\pi /4 \).

\section{Dynamical aspects of the PHE}

In the last section of our study of the PHE in ferrofluids, some dynamical
aspects are discussed, in relation to the rotation
of the magnetic moments of the particles
 induced by the magnetic field in ferrofluids.
In section \ref{Ferrofluids}, it was noted that the main process of
relaxation of the magnetic moments in the Cobalt samples is the brownian
mechanism. Since the Faraday effect is produced by the orientation
of the magnetic moments, it should follow the same
relaxation law as the magnetization. 
This relaxation of the Faraday rotation as a function of the frequency of the magnetic
field was indeed observed \cite{payet}. The characteristic relaxation time
can be obtained either from the study of the frequency dependence of the Faraday rotation
or from the study of the transient response as the magnetic field is turned off.
Both methods provide comparable values of the relaxation time. 
From Eq.~(\ref{tauB}), it can be noted that the relaxation time
for brownian relaxation is directly linked with the viscosity of the suspension. 
Using this idea, Bacri \emph{et al.} in Paris and B. Payet in St-Etienne
have experimentally shown that measurements of relaxation times 
could be used to determine the viscosity of the carrier liquid $\nu$. A viscosimeter
 based on this principle could achieve an accuracy of
a few percents and a range of several decades of viscosity. 
More recently, measurements of relaxation times are now used 
for biological applications of ferrofluids \cite{koetitz}. 

The magnetic susceptibility of ferrofluids is well described by Debye theory,
originally applied to dielectrics but also very useful for magnetic liquids.
Debye theory essentially relies on the following hypotheses:

\begin{enumerate}
\item All magnetic moments of the particles are fixed with respect to the particles
\item All particles have identical dimensions and  magnetic moments
\item There is no interaction between the particles. 
\end{enumerate}
Within Debye's model, the magnetic susceptibility is

\begin{equation}
\label{chi}
\chi (\omega _{H})=\frac{\chi _{0}}{1+i\omega _{H}\tau }=|\chi (\omega _{H}\tau )|e^{i\varphi (\omega _{H}\tau )},
\end{equation}
where \( \omega _{H} \) stands for the frequency of the magnetic field and \( \tau  \)
for the relaxation time of the liquid. 
The proportionality of the magnetic susceptibility (\ref{chi}) and the Faraday rotation
(measured both in \emph{amplitude} and \emph{phase}) has been established
 in Ref. \cite{payet}.
 Therefore, it is expected that the PHE should be also proportional to this
complex magnetic susceptibility. 
It is very important not to confuse the phase \( \varphi (\omega _{H}\tau ) \)
defined in Eq.~(\ref{chi}) with the phase of the scattered light, which
is often written as a complex-valued quantity.
The phase \( \varphi (\omega _{H}\tau ) \)
is defined unambiguously with respect to the magnetic field and varies at
the frequency of the magnetic field \( \omega _{H} \), whereas the phase of the
scattered light oscillates at optical frequency.

\begin{figure}
{\par\centering \resizebox*{8cm}{8cm}{\rotatebox{-90}{\includegraphics{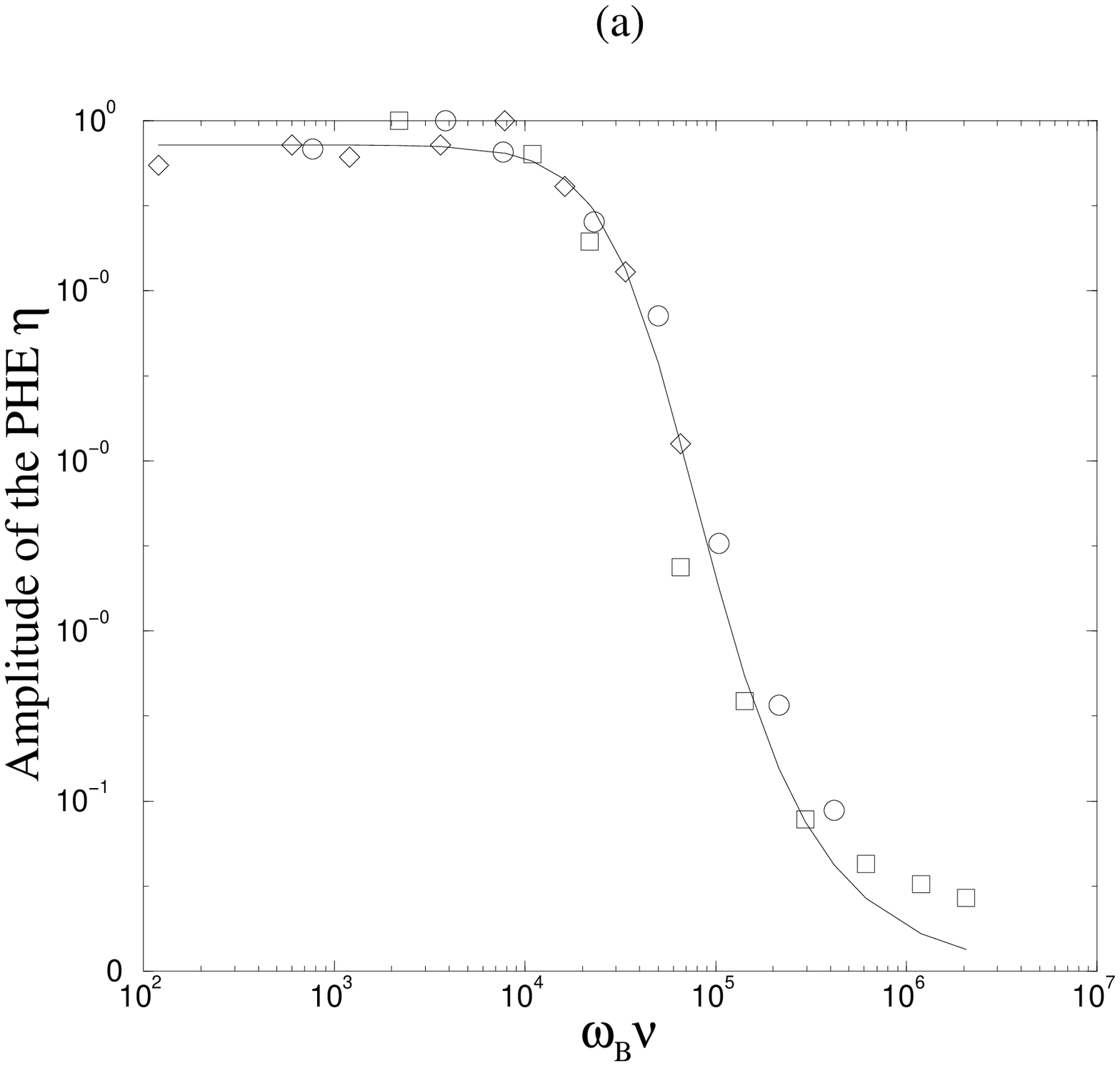}}} \par}
{\par\centering \resizebox*{8cm}{8cm}{\rotatebox{-90}{\includegraphics{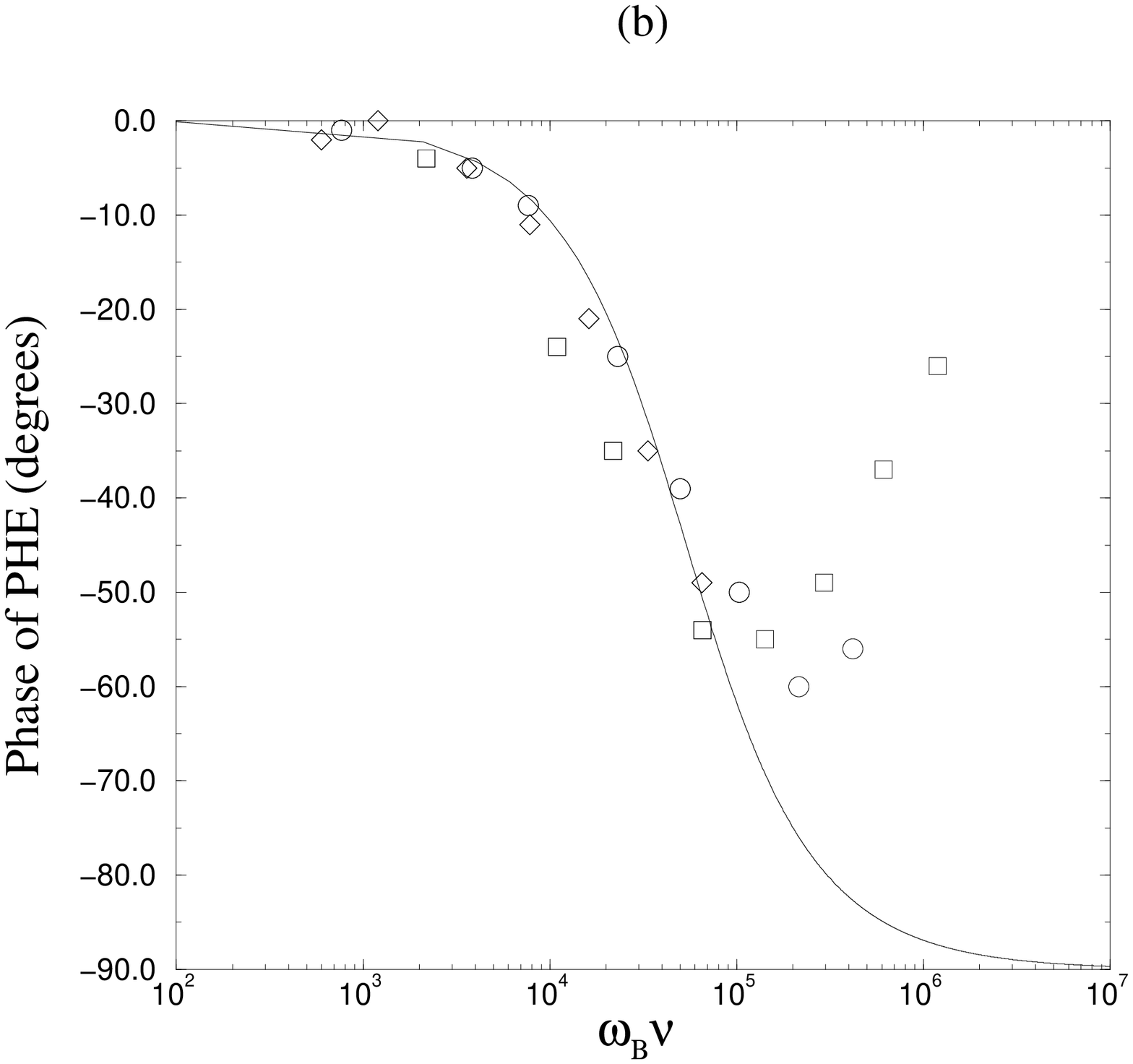}}} \par}
\caption{Amplitude (a) and phase (b) of normalized PHE as a function of the
product of the frequency of the magnetic field $\omega_B$ and the viscosity of the suspension
$\nu$ for samples of different viscosity: the original solution with \protect\( 100\protect \)\%
de glycerol (squares), a solution with \protect\( 90\protect \)\% of glycerol
(circles), and a solution with \protect\( 75\protect \)\% of glycerol (diamonds).
The solid curves in (a) and (b) are fits using Eq.~(\ref{chi}).
The experiment has been realized with a field of  \protect\( 100\protect \)Oe
and with samples of volume fraction  \protect\( 0.025\protect \)\%.
The Debye model predicts a behavior of the phase  
at high frequency, which is not seen in the experiment.
\label{Fig:freque}}
\end{figure}

In order to show the applicability of Debye's theory to the PHE in ferrofluids, 
 the amplitude and phase of the PHE were measured as function 
of the product \( \omega _{H}\tau  \),
as shown in Fig.~\ref{Fig:freque}.
Experiments were carried out in samples of
different viscosities by diluting a suspension of (liquid) ferrofluid with water, which
was originally a solution in glycerol. Three samples of identical volume fraction 
 \( f=0.025 \)\% in ferrofluid were used: the original solution with \( 100 \)\% of
 glycerol of viscosity \( 1093 \)Pa.s, a solution
with \( 90 \)\% of glycerol of viscosity \( 384 \)Pa.s, and a solution with \( 75 \)\%
of glycerol of viscosity \( 60 \)Pa.s at 25\( ^{\circ } \). 
A magnetic field of \( 100 \)Oe was applied during the experiment.
The curves for different viscosities 
do superimpose on each other,
which confirms that the PHE only depends on the product \( \omega _{H}\tau  \).

The saturation of the PHE as function of the magnetic field
and the measurement of relaxation time both confirm that
 the origin of the PHE resides in the orientation of
 the magnetic moments of the particles,
just like for the Faraday effect.
The observed behavior as function of frequency
 in figure \ref{Fig:freque}
is compared with the prediction of the Debye theory of Eq.~(\ref{chi}). 
The measurements of the amplitude and phase of the PHE agrees with the Debye model
when the parameter $\omega_B \nu$ is below a value of about $10^5$Pa.
Above this value, a deviation is seen between the experiment and the theoretical curve
 of the amplitude of the PHE, 
and a more dramatic deviation is seen in the phase at high frequency.
The experimental setup should not be responsible for this deviation, even at the 
 highest frequency of $\omega_B =1.875$kHz.
The PHE at this frequency is about two orders of magnitude
smaller as compared to its value at low frequency, but the
 noise level is still very low for these measurements (less than $1\%$).
In these rather dilute samples, the origin of this deviation, the caracteristic time of
 which is about $8$ms, is unclear.

\section{Conclusion}
We have presented a theoretical model of the PHE, which captures
the essential features of the experiments reported in this paper.
 This work confirms our understanding
 of the PHE in a system of Rayleigh absorbing scatterers, and its connection with
 the Faraday effect. In ferrofluids, as shown by static 
and dynamic measurements, the Faraday effect and the PHE
were both found to be produced by the orientation of the
 magnetic moments. 
This work on the PHE in ferrofluids could also be useful to improve
our understanding of complex media containing ferrofluid particles, such as media
in which the particles are not free or are only partially free to move.
The polarization dependence of the PHE 
was found to be sensitive to the degrees of freedom of 
the particles in the medium. In the magneto-transverse geometry, 
this type of experiment could provide information
 complementary to the one provided by the magnetic linear birefringence.\\
 
We wish to thank H. Roux, C. Bovier and G. Rikken for their help in the experimental part, 
J. Ferr\'e, R. Perzynsky, and J-C. Bacri for the enlightening discussions on ferrofluids,
 and T. Lubensky for stimulating discussions.


\newpage

\end{document}